\documentclass[letter]{aa}  
\usepackage{graphicx,bm,lineno}
\usepackage{txfonts}
\usepackage[colorlinks,citecolor=brown]{hyperref}
\usepackage{upgreek}
\usepackage{xspace}
\newcommand{\src}{WISEP\,J101905.63+652954.2\xspace}
\begin{document} 
   \title{Polarised radio pulsations from a new T~dwarf binary}
   \titlerunning{Radio pulsation from new T-dwarf binary}
   \authorrunning{Vedantham et al.}
   \author{H. K. Vedantham\inst{1,2}, Trent J. Dupuy\inst{3}, E. L. Evans\inst{3},  A. Sanghi\inst{4}, J. R. Callingham\inst{1,5}, T. W. Shimwell\inst{1,5}, W. M. J. Best\inst{5}, M. C. Liu\inst{6} and P. Zarka\inst{7}}
    \institute{
ASTRON, Netherlands Institute for Radio Astronomy, Oude Hoogeveensedijk 4, Dwingeloo, 7991 PD, The Netherlands
\and
Kapteyn Astronomical Institute, University of Groningen, PO Box 72, 97200 AB, Groningen, The Netherlands
\and
Institute for Astronomy, University of Edinburgh, Royal Observatory, Blackford Hill, Edinburgh, EH9 3HJ, UK
\and
The University of Texas at Austin, Department of Astronomy, 2515 Speedway, C1400, Austin, TX 78712, USA
\and
Leiden Observatory, Leiden University, PO Box 9513, 2300 RA, Leiden, The Netherlands
\and
Institute for Astronomy, University of Hawaii, 2680 Woodlawn Drive, Honolulu, HI 96822, USA
\and
LESIA, CNRS -- Observatoire de Paris, PSL 92190, Meudon, France
}
\date{Received XXX; accepted YYY}

\abstract{Brown dwarfs display Jupiter-like auroral phenomena such as magnetospheric H$\upalpha$ emission and coherent radio emission. 
Coherent radio emission is a probe of magnetospheric acceleration mechanisms and provides a direct measurement of the magnetic field strength at the emitter's location, both of which are difficult to access by other means. 
Observations of the coldest brown dwarfs (spectral types T and Y) are particularly interesting as their magnetospheric phenomena may be very similar to those in gas-giant exoplanets. 
Here we present 144\,MHz radio and infrared adaptive optics observations of the brown dwarf \src made using the LOFAR and Keck telescopes respectively. The radio data shows pulsed highly circularly polarised emission which yields a rotation rate of $0.32\pm0.03\,$hr$^{-1}$. The infrared imaging reveals the source to be a binary with a projected separation of $423.0\pm1.6$\,mas
between components of spectral type T5.$5\pm0.5$ and T7.$0\pm0.5$. 
%
%
With a simple ``toy model'' we show that the radio emission can in principle be powered by the interaction between the two dwarfs with a mass-loss rates of at least $25$ times the Jovian value.
\src is interesting because it is the first pulsed methane dwarf detected in a low radio-frequency search. Unlike previous gigahertz-frequency searches that were only sensitive to objects with kiloGauss fields, our low-frequency search is sensitive to surface magnetic fields of $\approx 50$\,Gauss and above which might reveal the coldest radio-loud objects down to planetary mass-scales.}

   \keywords{}

   \maketitle
\section{Introduction}
High energy charges around brown dwarfs are expected to be created by auroral (or magnetospheric) processes akin to that seen on gas-giant planets, as opposed to coronal and chromospheric acceleration expected on stars \citep{2012ApJ...760...59N,2018haex.bookE.171W,2017MNRAS.470.4274T}. This paradigm has been established based on highly circularly polarised and rotationally modulated radio pulses and H$\upalpha$ emission observed on brown dwarfs \citep{2007ApJ...663L..25H,2008ApJ...684..644H,2015Natur.523..568H,2012ApJ...747L..22R,2016ApJ...821L..21R,2017ApJ...834..117W}. The radio emission is of particular interest because it is expected to occur at the local cyclotron frequency, which in the non-relativistic limit, only depends on the ambient magnetic field strength \citep{1982ApJ...259..844M}. Because Zeeman splitting observations become very challenging in such cold objects as brown dwarfs due to the lack of non-broadened spectral lines, radio observations may be the only viable technique to directly measure their magnetic field strengths and topologies. In addition, unlike rocky planets, gas giants and brown dwarfs have predictable and relatively simple internal structures at depths where their magnetic field is expected to be generated \citep{2000ARA&A..38..337C}. This makes them ideal targets to test dynamo scaling laws \citep[e.g.,][]{2009Natur.457..167C} that are likely applicable even in the exoplanet regime. 

Despite concerted searches, radio detections of the coldest brown dwarfs are rare. The coldest, spectral type Y brown dwarfs have not been detected in the radio \citep{2019MNRAS.487.1994K}. At the warmer spectral type T, four brown dwarfs have been detected in dedicated surveys at radio frequencies of 5\,GHz and above \citep{2012ApJ...747L..22R,2016ApJ...818...24K,  2016ApJ...830...85R,2016ApJ...821L..21R,2016ApJ...818...24K,2018ApJS..237...25K}. Recently, we reported the first direct discovery of a brown dwarf made by virtue of its radio emission \citep{2020ApJ...903L..33V} using the LOFAR radio telescope \citep{2013A&A...556A...2V} at 144\,MHz. Here we report our second discovery also made with LOFAR. \src\ was originally discovered by \citet{2011ApJS..197...19K} in \textsl{Wide-field Infrared Survey Explorer} (\textsl{WISE}) data \citep{2010AJ....140.1868W} and, using spectroscopic data, assigned optical and near-infrared spectral types of T7 and T6, respectively.

Cold brown dwarfs share their radio phenomenology with Jupiter. The radio emission consists of two components. A quasi-quiescent component that is unpolarised or weakly polarised and a highly circularly polarised pulsed component that repeats at the rotation rate \citep{2018haex.bookE.171W,2013A&A...549A.131A,2008ApJ...684..644H,2006ApJ...648..629B}. However, the radio energetics of the detected brown dwarfs is orders of magnitude larger than that seen in Jupiter. This combined with a lack of detection of UV or H3+ from brown dwarfs \citep{2021A&A...655A..75S, 2022AJ....164...63G} suggest that the Jovian auroral energetics cannot be simply scaled to brown dwarfs. In any case, magnetic field lower limits derived from the pulsed component in three of the detected T~dwarfs have been over a factor of three larger than the predictions of leading dynamo scaling laws that can successfully predict the field strength of some solar system planets and low mass stars \citep{2018ApJS..237...25K}. This suggests that the model is inadequate, or it is also possible that by virtue of observing at high frequencies, the previous radio surveys were only sensitive to objects with anomalously large magnetic fields. For instance, \citet{2009Natur.457..167C} predict a field strength of $10^3$\,Gauss for a $50\,M_{\rm Jup}$ brown dwarf with an age of $10^9$\,yr and a surface temperature of $1500\,$K (late-L dwarf). The corresponding peak cyclotron frequency in its magnetosphere is about 2.8\,GHz which will make such an object undetectable in a 5\,GHz survey even if it were `typical' of the predicted population. The 144\,MHz LOFAR data can detect objects with surface field strengths as low as 50\,G. Therefore, the LOFAR-detected objects such as \src\ are beginning to provide a more complete sample to critically test dynamo scaling laws over a larger range in magnetic field strengths.

This paper is organised as follows: \S 2 presents details of the  radio and infrared observations and the analysis of the radio light curve. In \S 3 we discuss the possible mechanisms driving the radio emission, and present our conclusions and outlook in \S4. 

\section{Observations}
\subsection{LOFAR 144\,MHz observations}
\src\, was discovered as part of our ongoing search \citep[e.g.][]{2021NatAs...5.1233C} for stars, brown dwarfs, and exoplanets using data from the LOFAR Two Metre Sky Survey \citep[LoTSS;][]{2022A&A...659A...1S}. Our methodology has typically involved searching for circularly polarised sources in deep 8\,hr exposure LoTSS images. This is how we found {\em Elegast}, the first radio-selected brown dwarf \citep{2020ApJ...903L..33V}. Because brown dwarf auroral emission is typically pulsed at the rotation period, we have since implemented a search algorithm to construct Stokes-V light curves on various time-bin widths and search for on-off and periodic pulsations from known brown dwarfs.
Although we plan to conduct an untargeted search for such pulses over the Northern sky, we first validated our approach by a targeted search of ten known T- and Y-type brown dwarfs which led to the discovery of Stokes-V radio pulsations from \src.

\begin{figure*}
\centering
\includegraphics[width=0.95\linewidth]{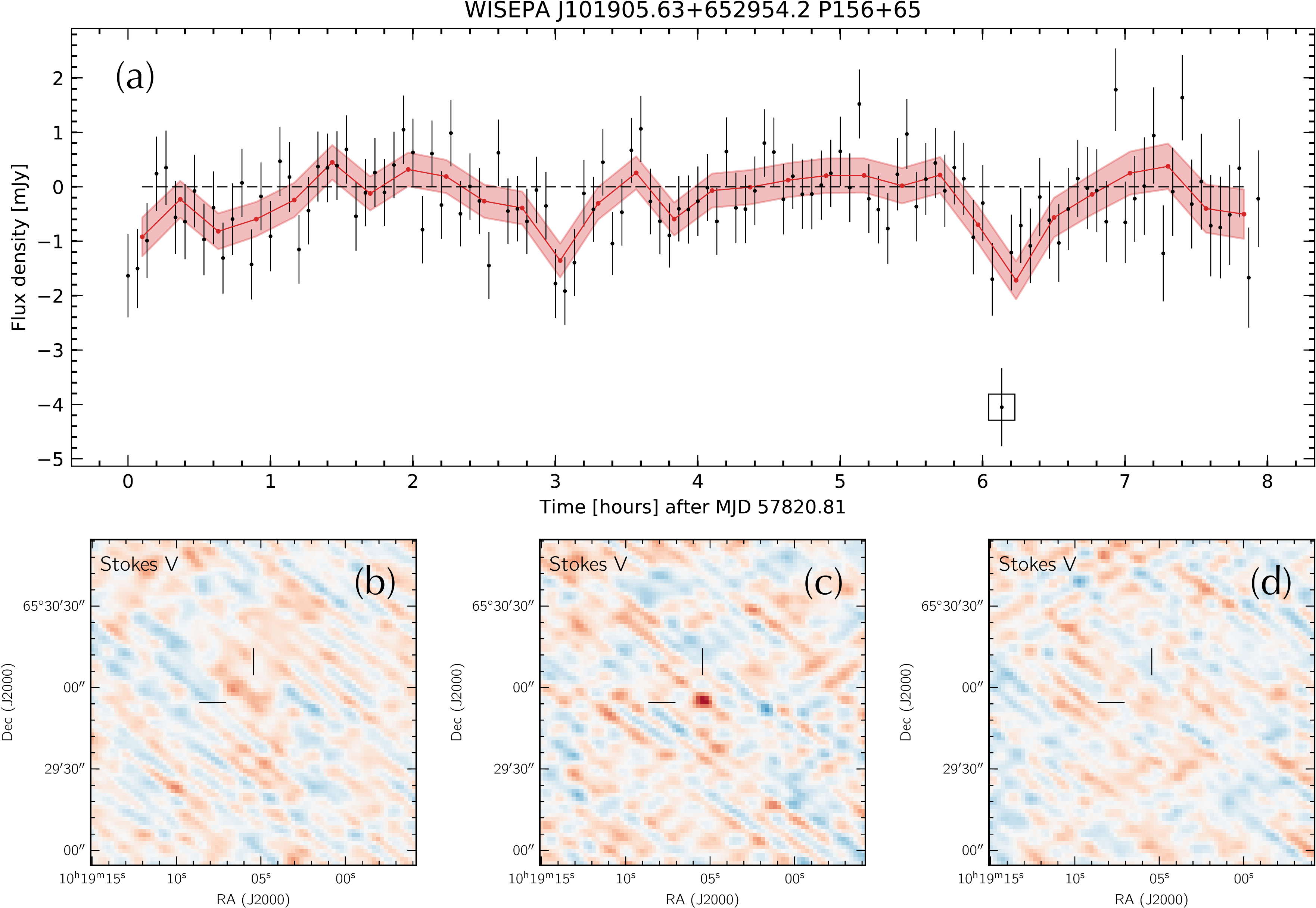}
\caption{Panel (a) shows the Stokes-V radio lightcurve of \src with a bin width of 4\,minutes (black points with $\pm 1\sigma$ errorbars) and 16\,minutes (red curve with shaded $\pm 1\sigma$ uncertainty). The point marked with the black square is a significant detection with a flux-density of $-4.1(7)\,{\rm mJy}$ whose Stokes-V image is shown in panel (c). Panels (b) and (d) show similar 4\,min exposure images bracketing the  integration show in panel (c).}
\label{fig:radio-discovery}
\end{figure*}

Our current pipeline takes in the standard calibrated visibilities from the LoTSS survey. We first subtract the LoTSS-detected sources from the visibilities using their direction dependent gains while retaining on-axis sources in the direction of the target for an additional round of self-calibration as described in \citet{2021A&A...651A.115V}. We then modelled and subtracted these sources using their clean components from \texttt{wsclean}. We then imaged the target fields using \texttt{wsclean} \citep{2014MNRAS.444..606O} at a cadence of 4\,minutes and extracted the light curves from these images after averaging over the available bandwidth. The light curve of \src shows a statistically significant burst (Fig. \ref{fig:radio-discovery}). The on and flanking off-burst snapshot images are also shown. The figure also shows the light curve binned to a resolution of 16\,min in red that reveals a hint of periodicity at around a 3\,hr period.  

Polarised radio emission from planets and brown dwarfs is expected to have a periodic signature at the rotation period of the object due to beaming (akin to a light house). To ascertain the period signature in the light curve, we computed the Lomb-Scargle periodogram of the light curve using the \texttt{astropy} \citep{2013A&A...558A..33A,2018AJ....156..123A} implementation (See Fig. 2). To compute the significance of the periodogram peaks we computed the false alarm rate based on the bootstrap method described in \citet{2018ApJS..236...16V}. We detect a dominant peak at a frequency of $0.324$\,hr$^{-1}$ with a false alarm rate of under 3\%. We compute an uncertainty in the peak's location of $0.033$\,hr$^{-1}$ using the method prescribed in equation 52 of \citet{2018ApJS..236...16V}.

\begin{figure}
    \centering
    \includegraphics[width=\linewidth]{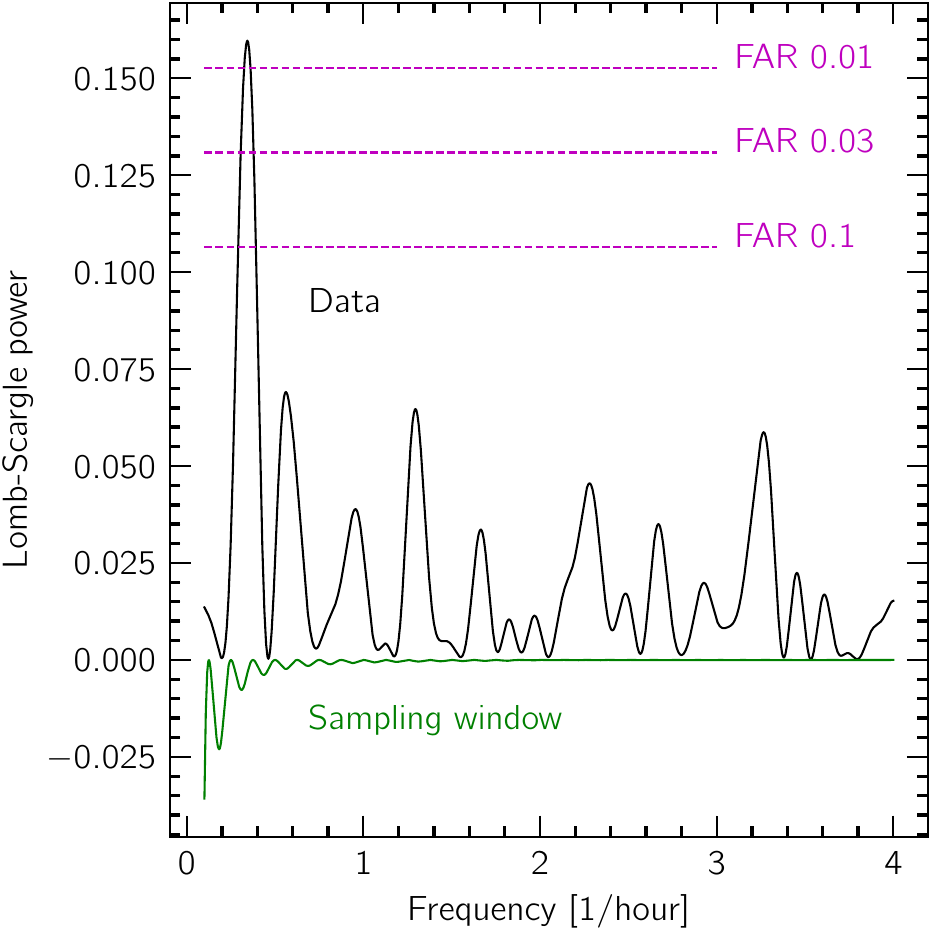}
    \caption{Lomb Scargle periodogram of the radio light curve from Fig. \ref{fig:radio-discovery}. The dominant peak with a false alarm rate of under 3\% is at $0.324\pm0.033$\,hr$^{-1}$.}
    \label{fig:my_label}
\end{figure}

\subsection{Keck/NIRC2 LGS AO} \label{sec:keck}

We observed \src on 2015 January 15\,UT and 2022 January 24\,UT with the facility imager NIRC2 at the Keck~II telescope in concert with the laser guide star (LGS) adaptive optics (AO) system \citep{2006PASP..118..310V,2006PASP..118..297W}. For tip-tilt correction, we used the star USNO-B1.0\,1554-0140735, which is 23\arcsec\ away from the target and provided flux to the tip-tilt sensor equivalent to $R=18.2$\,mag. The wavefront sensor monitoring the LGS measured flux equivalent to a $V=10.2$\,mag star in 2015 and $V=8.5$\,mag in 2022, thanks to the intervening LGS upgrade \citep{2016SPIE.9909E..0SC}. We obtained images with standard Maunakea Observatories filters in the $J$ and $H$ bands \citep{2002PASP..114..180T} as well as $CH_4s$, a medium-bandwidth filter centred on the $H$-band flux peak of T~dwarfs. For each filter, we obtained between four and six dithered 180-s images in 2015 and 60-s images in 2022 while keeping the LGS fixed to the centre of NIRC2's narrow camera (0\farcs01\,pixel$^{-1}$) field-of-view ($10\arcsec \times 10\arcsec$). In 2015, the AO correction deteriorated significantly as we collected data, and the quality of our $H$-band data set was too poor to be included in our analysis.

\begin{figure*}
    \centering
    \includegraphics[height=0.2\linewidth]{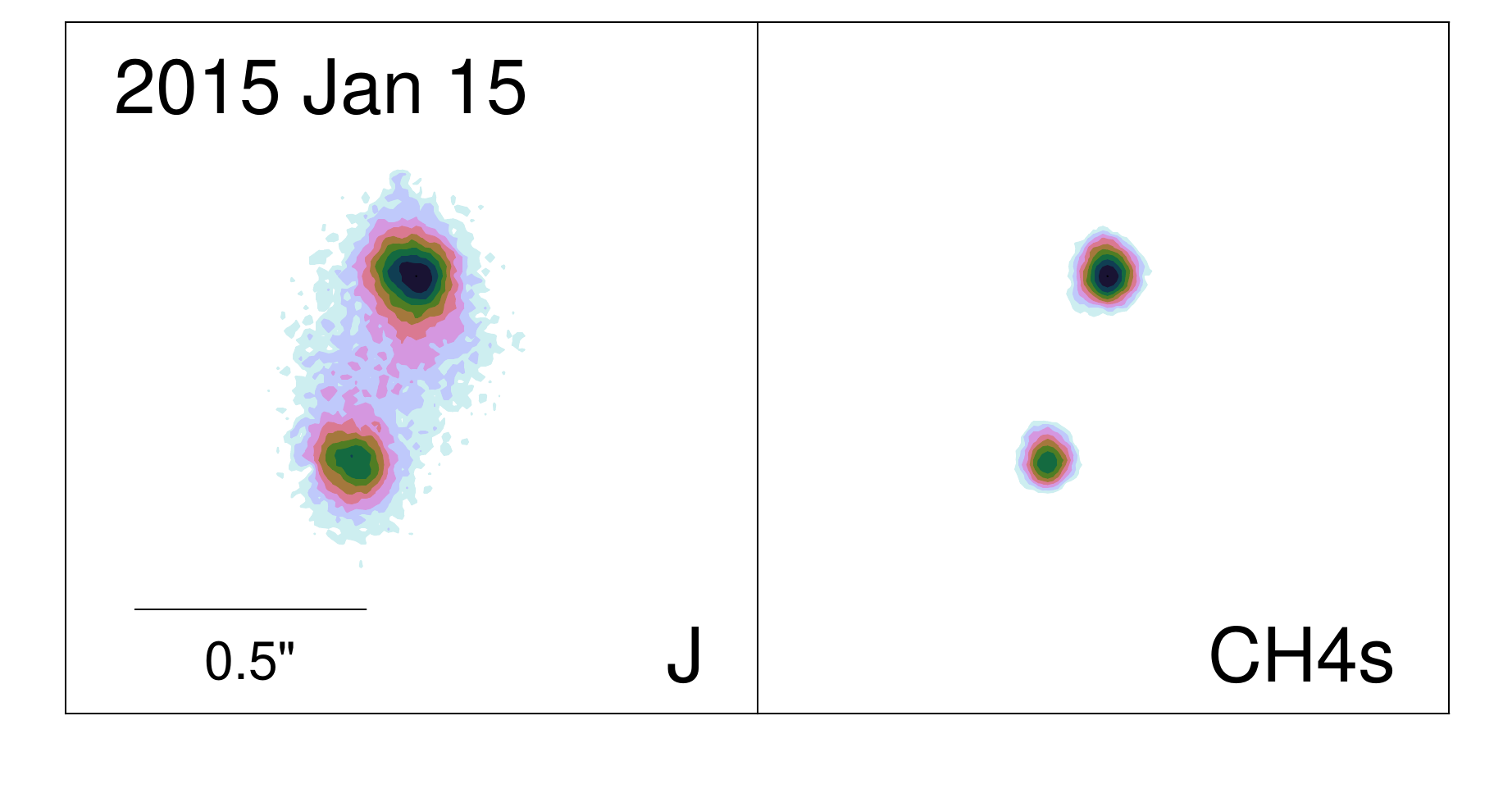}
    \includegraphics[height=0.2\linewidth]{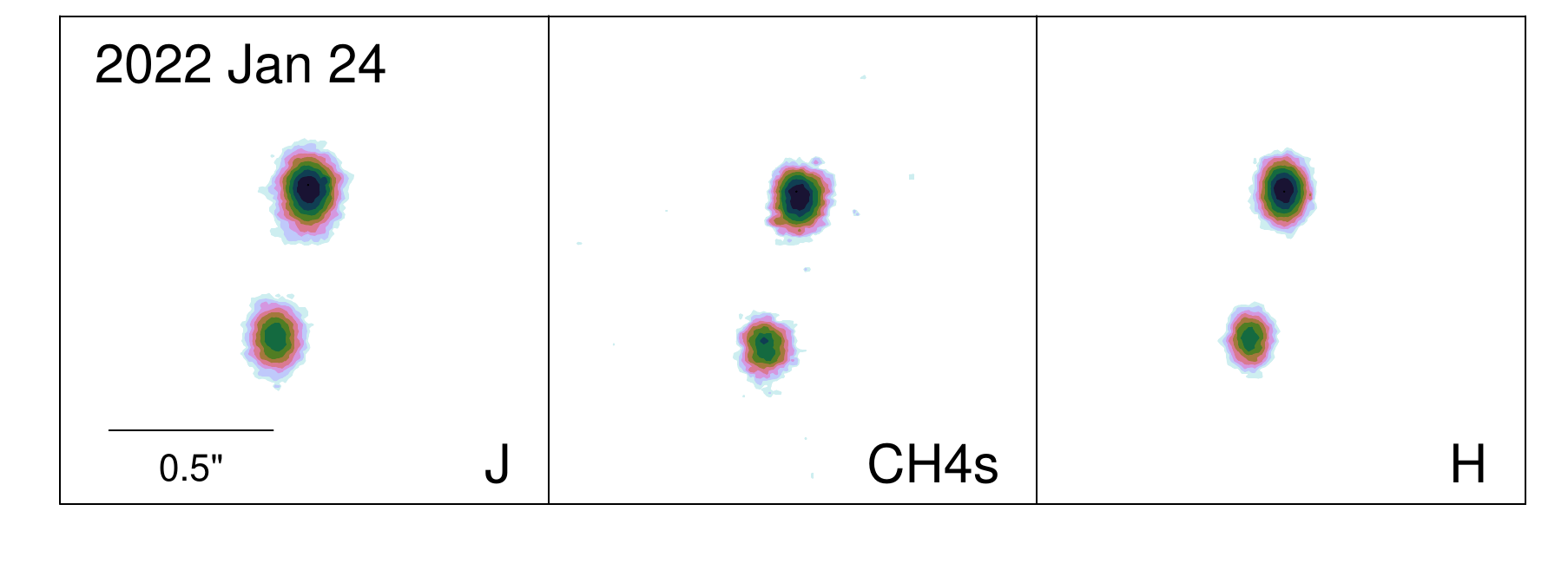}
    \caption{Contour plots of one typical individual exposure for each filter in which we obtained data. Contours are drawn in logarithmic intervals from the peak flux down to 10\% of the peak flux in each image. The images are all 1\farcs5 across with North up. In 2015, despite the AO correction deteriorating from 0\farcs09 in the $CH_4s$ band to 0\farcs13 in the $J$-band, the binary was still well resolved. We used the more precise differential magnitudes from the higher-quality, and fully contemporaneous 2022 images in our analysis.}
    \label{fig:binary}
\end{figure*}

We reduced our data using the same custom scripts as in our previous work \citep[e.g.,][]{2008ApJ...689..436L,2015ApJ...803..102D}, and examples of individual exposures are shown in Figure~\ref{fig:binary}. We measured the separation, position angle (PA), and magnitude difference in individual exposures using three-component, two-dimensional Gaussians, and computed the means and standard deviations of measurements from individual exposures as the final measurements and their uncertainties. For our 2015 data, we used the astrometric calibration of \citet{2010ApJ...725..331Y} to correct for nonlinear distortion, the orientation of NIRC2 (by subtracting $0\fdg252$), and the pixel scale ($9.952\pm0.002$\,mas\,pixel$^{-1}$). Likewise, for our 2022 data we used the calibration of \citet{2016PASP..128i5004S}. The resulting binary parameters are given in Table~\ref{tbl:keck}. Our relative astrometry is consistent within the errors at each epoch, and the repeated observations in $J$ and $CH_4s$ filters show no significant change in flux ratio. 

To compute the final relative astrometry at each epoch, we took the weighted average of our relative astrometry measurements and added a systematic error of 1.5\,mas to account for the uncertainty in the distortion corrections of NIRC2. This gives separations of $423.0\pm1.6$\,mas and $468.2\pm1.6$\,mas and PAs of $161\fdg71\pm0\fdg23$ and $166\fdg87\pm0\fdg20$, in 2015 and 2022, respectively. 
The observed motion of $\approx7$\,mas\,yr$^{-1}$ is much lower than the proper motion of the system ($150.6\pm1.1$\,mas\,yr$^{-1}$) measured by \citet{2019ApJS..240...19K}, so we conclude the companion shares a common proper motion with the primary and is physically bound.

\begin{table*}
\centering
\caption[]{Keck LGS AO Relative Astrometry and Photometry of \src.} \label{tbl:keck}
\begin{tabular}{lcccc}
\hline
Epoch (MJD) & Filter & Separation (mas) & PA ($\deg$) & $\Delta{m}$ (mag) \\
\hline
57037.5382 & $J$     & $416  \pm7  $ & $161.5 \pm0.6 $ & $0.37 \pm0.06 $ \\
57037.5246 & $CH_4s$ & $423.1\pm0.6$ & $161.72\pm0.12$ & $0.489\pm0.019$ \\
59603.5270 & $J$     & $467.2\pm1.1$ & $166.78\pm0.12$ & $0.494\pm0.021$ \\
59603.5218 & $CH_4s$ & $467  \pm3  $ & $166.82\pm0.18$ & $0.48 \pm0.03 $ \\
59603.5174 & $H$     & $468.7\pm0.7$ & $166.97\pm0.10$ & $0.579\pm0.013$ \\
\hline
\end{tabular}
\begin{list}{}{}
\item[Note.] Error bars given here are the standard deviation of individual measurements and do not account for the 1.5\,mas systematic error on the absolute astrometric reference frame of NIRC2 due to the optical distortion correction for such a wide binary. Relative photometry is given as the difference in magnitude $\Delta{m} \equiv m_{\rm B} - m_{\rm A}$.
\end{list}
\end{table*}

Our Keck images also showed a fainter point source $\approx$2\arcsec\ away from \src\ at a position angle of $\approx$200$^\circ$. We identified this source in the Pan-STARRS1 $3\pi$ Survey catalog \citep{2016arXiv161205560C} as PSO\,J154.7727+65.4978. It is visible in stacked $rizy_{\rm P1}$ images and appears brightest in $z_{\rm P1}$. Its $(z-y)_{\rm P1}=0.41\pm0.13$~mag color (using stacked photometry) is far too blue to be a fainter, later-T or Y dwarf \citep{2018ApJS..234....1B}, so we conclude this is a background star or galaxy.
Although this source is only about 2\arcsec\ from the nominal position of the radio detection, it is almost certainly not the source of the observed radio emission. The high circular polarisation in the radio-band is inconsistent with an extragalactic origin, so we only need consider the Galactic stellar hypothesis. The absolute radio astrometry has a Gaussian-equivalent standard deviation of $\sigma \approx 0\farcs5$  \citep{2022A&A...659A...1S} yielding a $4\sigma$ discrepancy in position. The Pan-STARRS1 $z-y$ colour suggests that the source is a mid M-dwarf whose $z_{\rm P1}=21.06\pm0.06$\,mag places it at a photometric distance of over 300\,pc. This is well beyond the sensitivity horizon of LOFAR for M-dwarfs' cyclotron maser emission \citep{2021NatAs...5.1233C}. Finally, the rotation rate implied by the radio observations of  $0.32\,{\rm hr}^{-1}$ is unusually large for a mid M-dwarf \citep{2021ApJ...916...77P}. For these reasons, we reject the association between the radio source and PSO\,J154.7727+65.4978.

In order to compute $CH_4s-H$ colors for the two components of \src from Keck~LGS~AO imaging, we used its IRTF/SpeX spectrum from 2010~May~27\,UT \citep{2011ApJS..197...19K} to measure integrated-light colors of $CH_4s-H = -0.42$\,mag and $J-H = -0.34$\,mag. Combined with the 2MASS measurement of $J=16.589\pm0.055$\,mag, these colors give integrated-light photometry of $H=16.93\pm0.06$\,mag and $CH_4s=16.51\pm0.06$\,mag. Combined with our measured magnitude differences in $CH_4s$ and $H$, we find colors of $CH_4s-H=-0.382\pm0.013$\,mag and $-0.481\pm0.020$\,mag for the primary and secondary. Using the spectral type-colour relation detailed by \citet{2008ApJ...689..436L}, we determine methane-photometry-based spectral types of T$5.5\pm0.5$ and T$7.0\pm0.5$.

\section{Discussion}
\subsection{Mass and magnetic field estimates} \label{sec:mass}
We computed the combined-light bolometric luminosity of \src by direct integration of its unresolved optical to mid-infrared (MIR) spectral energy distribution (SED). The assembled SED consists of available Pan-STARRS-1 \citep[PS1;][]{2016arXiv161205560C} optical photometry ($z$, $y$), the near-infrared (NIR) IRTF/SpeX prism spectrum, NIR photometry from 2MASS \citep{2003tmc..book.....C} and MKO \citep{2021AJ....161...42B}, and MIR photometry from the CatWISE catalog \citep[$W1$ and $W2$ bands;][]{2020ApJS..247...69E,2021ApJS..253....8M}, AllWISE catalog \citep[$W3$ and $W4$ bands;][]{2013wise.rept....1C}, and Spitzer/IRAC Channels 1 and 2 \citep{2004ApJS..154...10F}. First, we flux-calibrated the SpeX spectrum using the weighted average of scale factors derived from PS1 $y$, 2MASS $JHKs$, and MKO $JHK$ photometry, assuming a systematic noise floor of 0.01 mag for all the filters. We then integrated the flux-calibrated SpeX spectrum to determine the NIR contribution to the bolometric flux, accounting for the uncertainties in the spectral data points and the flux calibration procedures. We determined the optical and MIR contributions to the bolometric flux by simultaneously fitting ATMO model atmospheres \citep{2020A&A...637A..38P} to the PS1 and WISE photometry (computing synthetic photometry from the models) and the SpeX spectrum (with the models degraded to the non-linear spectral resolution of the 0$\farcs$5 slit). We found the best-fitting ATMO model had $T_{\mathrm{eff}} = 1000$ K and log $g = 5.5$ dex. Our final bolometric flux was found by adding the NIR contribution to the integration of the model outside the wavelength range of the SpeX spectrum. The uncertainty in the optical+MIR contribution was obtained from the standard deviation of the corresponding measurements derived using the three model spectra adjacent in $T_{\mathrm{eff}}$ and log $g$ to the best-fitting model. Using WISEPA J101905.63+652954.2’s parallax of $42.9 \pm 1.8$\,mas, we calculated its bolometric luminosity $\mathrm{log}(L_{\mathrm{bol}}/L_\odot) = -4.994 \pm 0.063$ dex.

To determine the mass of each component in the binary system, we combined their luminosities and ages with the \citet[]{2008ApJ...689.1327S} (SM08) hybrid evolutionary model. Each component’s luminosity is estimated using the bolometric luminosity vs spectral type relations from \citet{2015ApJ...810..158F}. We find that $\approx$70\% of the luminosity is contributed by the T5.5 dwarf and $\approx$30\% is contributed by the T7 dwarf. For each object, we adopt the field-age distribution from \citet[DL17][]{2017ApJS..231...15D}. For our mass calculations, we use the Bayesian rejection sampling technique described in \citet{2017ApJS..231...15D}. First, we draw $10^6$ random (luminosity, age) samples from a uniform distribution spanning the bolometric luminosity range of the evolutionary model grid and the intersection of the DL17 age range and the evolutionary model grid age range. Second, we compute the probability of each sample based on the $\chi^2$ of the drawn luminosity with respect to the measured value and the likelihood of drawing the sample’s age from the DL17 distribution. Third, we randomly draw $10^6$ uniform variates ($u$) distributed in the range from 0 to 1 and reject any samples where $p < u$. The fourth and final step is to linearly interpolate the evolutionary models (in logarithmic space) at each accepted luminosity-age point to calculate the corresponding mass. We find a mass of $41 \pm 18 \; M_{\mathrm{Jup}}$ for the T5.5 component and $32 \pm 16 \; M_{\mathrm{Jup}}$ for the T7 component.

Armed with the mass and luminosity values, we can estimate the magnetic fields of the two objects using so-called dynamo scaling laws. We employed the `saturated dynamo' scaling law proposed by \citet{2009Natur.457..167C} that relates the magnetic field to the heat flux and mean density of the brown dwarf. We used the law in the form given by \citet[][ their equation 1]{2010A&A...522A..13R}. We also used their correction to estimate the surface dipolar field from the field at the top of the dynamo as predicted by the scaling law \citep[][ their equation  2]{2010A&A...522A..13R}.  Although the objects' luminosities are have small errors, the mass estimates and the normalising constant in the scaling law have large fractional errors. To properly incorporate these errors into the predicted magnetic field strength, we ran a Monte-Carlo simulation where we drew the normalising constant from a uniform distribution \citep[][their equation 1]{2009ApJ...697..373R}, and the mass from a normal distribution. In each step of the Monte Carlo run we interpolated the evolutionary models of \citet{2003A&A...402..701B} to find the relationship between mass and field strength for the measured luminosity (i.e. for different ages). The resulting distribution of polar dipole field strengths for the T5.5 and T7 objects had a mean and standard deviation of $660\pm300\,$G and $460\pm210\,$G respectively. 

The observed cyclotron maser emission itself places a lower limit on the polar surface magnetic field strength of $51.4\,$G (cyclotron frequency at the mid-point of the LOFAR data's radio band). While this is consistent with the field estimates made above, higher frequency observations are necessary to critically test the dynamo scaling law. In what follows, we will leave the polar surface field as a variable while normalising our equations at $B=1\,{\rm kG}$.

\subsection{Energetics}
\src\,has not yet been detected at the gigahertz-frequencies where quiescent incoherent synchrotron emission is typically observed. A non-detection in the VLA Sky Survey \citep{2020PASP..132c5001L} quick-look images yields a $3\sigma$ upper limit of $0.34\,$mJy in the 2--4\,GHz band. Although incoherent radio emission has widely been used as proxy for the energetics of magnetospheric and coronal emitters \citep{2017ApJ...846...75P,2021MNRAS.507.1979L,1994A&A...285..621B}, here we use the pulsed radio emission to calculate the energetics of the auroral electrons. We posit that the radio pulsations are due to beaming combined with rotation and that the beam solid angle of the radio emission is 1.6\,sr --- identical to that of Jupiter's auroral radio emission due to its magnetosphere--ionosphere coupling \citep{2004JGRA..109.9S15Z}. 
The radio spectral luminosity for a pulse flux density of 2\,mJy (see Fig. \ref{fig:radio-discovery}) and a measured distance of 23.3\,pc \citep{2019ApJS..240...19K} is then  $(2\,{\rm mJy})\times (23.3\,{\rm pc})^2 \times (1.6\,{\rm sr}) \approx 1.7\times 10^{14}\,{\rm erg}\,{\rm s}^{-1}\,{\rm Hz}^{-1}$. Let us further assume that the total bandwidth of the radio emission is equal to the cyclotron frequency at the surface of the object. Then the auroral radio power is $4.6\times 10^{23}\,(B/{\rm kG})\,{\rm erg}\,{\rm s}^{-1}$. Assuming a 1\% efficiency in the conversion of the available auroral power to radio waves \citep{2007P&SS...55..598Z,2011JGRA..116.4212L}, we obtain an auroral power of $4.6\times 10^{25}\,(B/{\rm kG})\,{\rm erg}\,{\rm s}^{-1}$. For comparison, the auroral power output of Jupiter is $\sim 10^{20}\,{\rm erg}\,{\rm s}^{-1}$ \citep{2000RvGeo..38..295B}, and \citet{2017MNRAS.470.4274T} predict auroral powers of up to $10^{26.5} \,{\rm erg}\,{\rm s}^{-1}$ (assuming the same 1\% radio efficiency) for the Jovian magnetosphere-ionosphere paradigm applied to ultra-cool dwarfs. 

\subsection{Is binary interaction powering the radio emission?}
Magnetic interaction between the two objects can accelerate charges that eventually emit cyclotron maser radio emission, as seen in the Jupiter-Io system \citep{1969ApJ...156...59G,1998JGR...10319843N,1998JGR...10320159Z}. 
The projected separation of the two brown dwarfs in \src is $9.9\pm0.4\,$au, given their parallactic distance of $23.3\pm1.0$\,pc \citep{2019ApJS..240...19K}. 
We explored the full range of orbital parameters for the binary by fitting the two epochs of relative astrometry from Section~\ref{sec:keck} with the {\sc orvara} orbit analysis tool \citep{2021AJ....162..186B}. We used a prior on the total mass of $0.071\pm0.033\,M_{\odot}$ based on our mass estimates from Section~\ref{sec:mass}. As expected, the orbital parameters are poorly constrained, but we can place 3$\sigma$ limits on the semimajor axis ($>5.2$\,au), period ($>30$\,yr), inclination ($<86^{\circ}$), and eccentricity ($<0.96$). The posterior distributions have medians and 1$\sigma$ confidence intervals of $11^{+4}_{-6}$\,au, $160^{+120}_{-130}$\,yr, and $69^{+13}_{-11}\,{\rm deg}$, but we caution that these are highly influenced by the priors. (The eccentricity posterior is almost unchanged from the uniform prior below the upper limit we quote.)

Based on the radio rotation rate of the emitter, its light cylinder is at a radial distance of about $3.4\,$au. Therefore, even if the magnetospheres are not loaded with plasma (i.e. under force-free electrodynamics), direct magnetic interaction between the two dipolar magnetospheres is not possible and we must consider interception by one brown dwarf of the Poynting flux radiated by the other. The Poynting flux radiated by an oblique rotator (akin to a Pulsar's dipole emission) is of the order $L \sim B_0^2R_0^6\Omega^4/c^3$ \citep{2016era..book.....C} where $B_0$ is the surface magnetic field, $\Omega$ is the angular rotation rate, and $R_0$ is the object's radius. For characteristic values of $B_0 = 10^3\,{\rm G}$, $R_0 = 7\times 10^9\,{\rm cm}$, and $\Omega = 5.6\times 10^{-4}\,{\rm s}^{-1}$, we get $L\sim 10^{20}\,{\rm erg}\,{\rm s}^{-1}$ which falls well short of the value necessary to power the radio emission.  

Next, consider a scenario where the magnetospheres are loaded by plasma and drive a feeble wind. For simplicity, let us assume that the two magnetospheres and their co-rotation rates are similar. Due to the fast rotation, the balance between the centrifugal force of the co-rotating plasma and magnetic pressure must determine the structure of the magnetosphere in this case (i.e. gravitational force can be safely neglected) and the eventual Poynting flux. The centrifugal pressure felt by the plasma is $F_{\rm c} = \rho\Omega^2R^2/2$ where $R$ is the radial distance, $\Omega$ is the angular rotation rate and $\rho = \rho(R)$ is the plasma density at radius $R$. The magnetic pressure for a dipole at distance $R$ is $F_{B} = B_0^2R^{-6}R_0^6/(8\pi)$ where $R_0$ is the object's radius and $B_0$ is the surface magnetic field strength. In our simple `toy model', at low radii, $F_B$ dominates enforcing co-rotation with a dipolar field. This breaks at a critical radius when $F_{B} = F_{\rm C}$. Beyond this radius, we assume that the field lines open up into a Parker-spiral type configuration.  Note that $F_B=F_{\rm C}$ is equivalent to saying that the co-rotation speed equals the local Alfv\'{e}n speed. The critical radius is therefore the so-called Alfv\'{e}n point:
\begin{equation}
    r_A = \left(\frac{B_0^2R_0^6}{4\pi \Omega^2 \rho(r_A)} \right)^{1/8},
\end{equation}
In the open field zone, the azimuthal field dominates, falling off with distance, $R$ as $R^{-1}$. We therefore assume $B(R) = B(r_A)(R/r_A)^{-1}$ where $B(r_A) = B_0(r_A/R_0)^{-3}$. The brown dwarf wind beyond $r_A$ is assumed to  to have a radial flow speed, $v_r$ equal to the co-rotation speed at $r_A$ as suggested for the Jovian case by \citet{1974GeoRL...1....3H}. With these assumptions, the Poynting luminosity can be readily computed as $S = (B^2/8\pi)\times v_r\times (4\pi R^2)$ at any closed surface of radius $R>r_A$. The mass-loss rate is given by  $\dot{M} = (4\pi r_A^2)\times v_r \times \rho(r_A)$. For parameters applicable to \src of $R_0=7\times 10^9\,{\rm cm}$, $B_0=10^3\,{\rm G}$, $\Omega = 5.6\times 10^{-4}\,{\rm s}^{-1}$, we find that the necessary Poynting luminosity of $\approx 10^{25.5}\,{\rm erg}\,{\rm s}^{-1}$ can be achieved with a mass-loss rate of $\approx 25$\,tonnes per second. The corresponding Alfv\'{e}n point is at $r_A = 188R_0$. If instead we assume $B_0=100\,{\rm G}$ then we get the necessary Poynting flux for $\dot{M} \approx 550\,$tonnes per second and $r_A = 40R_0.$ For comparison, Io's volcanism is the principal source of Jovian magnetospheric plasma whose loss rate is about 1\,tonne per second. 
In any case, a significant fraction of the emitted Poynting flux must be intercepted by the magnetosphere of the companion for conversion of this Poynting flux into radiation emission due to binary interaction. We therefore conclude that while energetically feasible in principle, further work on the precise details of the wind--wind interaction and the source of mass-loss must be worked out to ascertain whether this interaction could have powered the observed radio emission from \src. 

\subsection{Auroral signatures}
Regardless of the veracity of the interaction-powered emission scenario, let us assume that at the emitter in \src, an auroral mechanism similar to that seen on Jupiter is at play. Such aurorae have also been suggested as the radio emission mechanism in other brown dwarfs and ultracool dwarfs \citep[e.g.][]{2015Natur.523..568H,2017MNRAS.470.4274T}.
Jupiter's aurorae emit comparable amounts of power in the radio and H$\upalpha$ line \citep{2000RvGeo..38..295B,1998JGR...10320159Z}. Assuming the same for \src, we would anticipate an H$\upalpha$ luminosity of $4.6\times 10^{23}\,(B/{\rm kG})\,{\rm erg}\,{\rm s}^{-1}$. Assuming a characteristic line width of $6\AA$ \citep{2016ApJ...826...73P}, the expected H$\upalpha$ flux density is $\approx 7\times 10^{-18}\,(B/{\rm kG})\,{\rm erg}\,{\rm s}^{-1}\,{\rm cm}^{-2}\,\AA^{-1}$. Based on the optical spectrum or \src presented by \citet{2011ApJS..197...19K}, we derive a $2\sigma$ upper limit on the H$\upalpha$ luminosity of $2.8\times 10^{-18}\,{\rm erg}\,{\rm s}^{-1}\,{\rm cm}^{-2}\,\AA^{-1}$. This suggests that the surface magnetic field of \src\,is $B\lesssim 10^3\,{\rm G}$ which is broadly consistent with our magnetic field estimate from \S3.1. Nevertheless, we caution that it is not possible to make definite statements on the magnetic field strength because the radio and H$\upalpha$ efficiencies and the radio beam solid angle can only be trusted to within an order of magnitude. In conclusion, we find that the available data are consistent with a Jupiter-like auroral process driving the radio emission in a magnetosphere with a surface strength of order ap kiloGauss. 

\section{Conclusions \& Outlook}
Magnetospheric emissions from the coldest brown dwarfs provide a rare glimpse into magnetism in the planetary mass regime outside the solar system. 
Here we have presented our second detection of a methane-bearing, T-type brown dwarf---\src---with LOFAR at 144\,MHz. The radio emission is pulsed and periodic, from which we derive a rotation rate of $0.32\pm0.03\,{\rm hr}^{-1}$ ($1\sigma$ bounds). We have also presented infrared adaptive optics observations of \src that show it to be a T-dwarf binary with a separation of $9.9\pm0.4$\,au and spectral types T$5.5\pm0.5$ and T$7.0\pm0.5$, making it the first T-dwarf binary to be detected in the radio band. We considered binarity as the cause of the radio emission. We find that while energetically feasible for mass-loss rates of $\gtrsim 25$\,tonnes per second, precise details of the interaction region must be studied before binary-interaction can be posited as the probably cause of the emission. In this regard, it is interesting to note that \citet{2022ApJ...932...21K} have suggested (based on detection rates and luminosities) that binary ultracool dwarfs may be more radio-loud than their single counterparts. If this is true, then a radio-selection as we have done here might reveal a population of close binary brown dwarfs upon infrared follow-up observations, similar to \src. 

\src is the first brown dwarf detected at 144\,MHz with the canonical periodic pulsed emission profile similar to that seen in the cm-wave band and on Jupiter at $\nu\lesssim 40 \,{\rm MHz}$. Three previously detected T-dwarfs in the cm-wave band have, unexpectedly, shown pulses up to 10 and/or 15 GHz with no sign of a distinct high-frequency cut off \citep{2018ApJS..237...25K}. This suggests magnetic field strengths well in excess of that anticipated by some dynamo scaling laws suggesting that the laws need to be revised. However, it is also possible that by virtue of a survey bias, the high frequency surveys have preferentially detected a small population of T~dwarfs that have anomalously high field strengths possibly in smaller magnetic loops rather than the large scale field predictions made from dynamo models. Because \src was selected from a 144\,MHz survey that does not have this selection bias, it will be very interesting to see if its spectral cut-of continues to unexpected trend discovered by \citet{2018ApJS..237...25K}.

We end by noting that \src is the second detected, and first pulsed, brown dwarf system found in the ongoing LOFAR Two Metre Sky Survey. As demonstrated by \citet{2020ApJ...903L..33V}, because the radio emission is non-thermal in origin, radio surveys may be able to discover a population of the coldest brown dwarfs that are too faint to be found in canonical infrared surveys. The pulsed emission from \src therefore motivates an all-sky, untargeted search for pulsed, circularly-polarised emitters in LoTTS survey data. 

\bibliographystyle{aa}
\bibliography{ref}

\begin{acknowledgements}
We thank Dr. Davy Kirkpatrick for making the Keck optical spectrum of \src available to us in machine-readable format.
HKV acknowledges funding from the Dutch Research Council (NWO) for the project e-MAPS (project number Vi.Vidi.203.093) under the NWO talent scheme VIDI. 
JRC thanks NWO for support via the Talent Programme Veni grant. 
LOFAR is the Low Frequency Array designed and constructed by ASTRON. It has observing, data processing, and data storage facilities in several countries, that are owned by various parties (each with their own funding sources), and that are collectively operated by the ILT foundation under a joint scientific policy. The ILT resources have benefitted from the following recent major funding sources: CNRS-INSU, Observatoire de Paris and Universit\'{e} d'Orl\'{e}ans, France; BMBF, MIWF-NRW, MPG, Germany; Science Foundation Ireland (SFI), Department of Business, Enterprise and Innovation (DBEI), Ireland; NWO, The Netherlands; The Science and Technology Facilities Council, UK. This research made use of the Dutch national e-infrastructure with the support of the SURF Cooperative (e-infra 180169) and the LOFAR e-infra group. The J\"{u}lich LOFAR Long Term Archive and the German LOFAR network are both coordinated and operated by the J\"{u}lich Supercomputing Centre (JSC), and computing resources on the supercomputer JUWELS at JSC were provided by the Gauss Centre for Supercomputing e.V. (grant CHTB00) through the John von Neumann Institute for Computing (NIC). This research made use of the University of Hertfordshire high-performance computing facility and the LOFAR-UK computing facility located at the University of Hertfordshire and supported by STFC [ST/P000096/1], and of the Italian LOFAR IT computing infrastructure supported and operated by INAF, and by the Physics Department of Turin University (under an agreement with Consorzio Interuniversitario per la Fisica Spaziale) at the C3S Supercomputing Centre, Italy. 
Some of The data presented herein were obtained at the W. M. Keck Observatory, which is operated as a scientific partnership among the California Institute of Technology, the University of California and the National Aeronautics and Space Administration. The Observatory was made possible by the generous financial support of the W. M. Keck Foundation. The authors wish to recognise and acknowledge the very significant cultural role and reverence that the summit of Maunakea has always had within the indigenous Hawaiian community.  We are most fortunate to have the opportunity to conduct observations from this mountain.
For the purpose of open access, the author has applied a Creative Commons Attribution (CC BY) licence to any Author Accepted Manuscript version arising from this submission.
\end{acknowledgements}

\end{document}